%% file: paper.tex
\newcommand*{\EXTENDEDVERSION}{}

\ifdefined\EXTENDEDVERSION
\documentclass[sigconf,svgnames,table,anonymous=false]{acmart}
\else
\documentclass[sigconf,svgnames,table,review,anonymous]{acmart}
\fi

\setcopyright{rightsretained}

\ifdefined\EXTENDEDVERSION
\renewcommand\footnotetextcopyrightpermission[1]{}
\fi

\ifdefined\EXTENDEDVERSION
\else
\acmConference[ESEC/FSE 2018]{The 26th ACM Joint European Software Engineering Conference and Symposium on the Foundations of Software Engineering}{4--9 November, 2018}{Lake Buena Vista, Florida, United States}
\acmYear{2018}
\acmISBN{} % \acmISBN{978-x-xxxx-xxxx-x/YY/MM}
\acmDOI{} % \acmDOI{10.1145/nnnnnnn.nnnnnnn}
\fi

\startPage{1}

\setcopyright{none}
\usepackage{booktabs}   %% For formal tables:
\usepackage{subcaption} %% For complex figures with subfigures/subcaptions
\usepackage{xcolor}
\usepackage{pgfpages}
\usepackage{pgfplots}
\usepackage{tikz}
\usepackage{enumerate}
\usepackage{algorithmicx}
\usepackage{algorithm}
\usepackage[noend]{algpseudocode}
\usepackage{graphicx}
\usepackage{xspace}
\usepackage[super]{nth}
\usepackage{multirow}
\usepackage{flushend}

\definecolor{light-gray}{gray}{0.5}
\input{solidity-highlighting.tex}
\usetikzlibrary{automata,calc,trees,positioning,arrows,chains,shapes.geometric,%
  decorations.pathreplacing,decorations.pathmorphing,decorations.text,shapes,%
  matrix,shapes.symbols,patterns,shadows,arrows.meta}
\pgfplotsset{compat=newest}
\newcommand*\Let[2]{\State #1 $\gets$ #2}
\newcommand*\Fcall[1]{\textsc{#1}}
\algrenewcommand\alglinenumber[1]{\tiny\gray{#1}}
\algrenewcommand\algorithmicforall[2]{\textbf{foreach} #1 \textbf{in} #2}
\algrenewtext{ForAll}{\algorithmicforall}
\algnewcommand\algorithmicswitch{\textbf{switch}}
\algnewcommand\algorithmiccase{\textbf{case}}
\algdef{SE}[SWITCH]{Switch}{EndSwitch}[1]{\algorithmicswitch\ #1\ \algorithmicdo}{\algorithmicend\ \algorithmicswitch}%
\algdef{SE}[CASE]{Case}{EndCase}[1]{\algorithmiccase\ #1}{\algorithmicend\ \algorithmiccase}%
\algtext*{EndSwitch}%
\algtext*{EndCase}%
\newcommand\gray[1]{\textcolor{light-gray}{#1}}
\newcommand\code[1]{\lstinline[language=Solidity,basicstyle=\small\ttfamily]{#1}\xspace}

\newcommand\secref[1]{Sect.~\ref{#1}}
\begin{document}

%% Title information
\title[Specification Mining for Smart Contracts]
      {Specification Mining for Smart Contracts\\ with Automatic Abstraction Tuning}
                                        %% [Short Title] is optional;
                                        %% when present, will be used in
                                        %% header instead of Full Title.
%% \titlenote{with title note}             %% \titlenote is optional;
%%                                         %% can be repeated if necessary;
%%                                         %% contents suppressed with 'anonymous'
%% \subtitle{Subtitle}                     %% \subtitle is optional
%% \subtitlenote{with subtitle note}       %% \subtitlenote is optional;
%%                                         %% can be repeated if necessary;
%%                                         %% contents suppressed with 'anonymous'

%% Author information
%% Contents and number of authors suppressed with 'anonymous'.
%% Each author should be introduced by \author, followed by
%% \authornote (optional), \orcid (optional), \affiliation, and
%% \email.
%% An author may have multiple affiliations and/or emails; repeat the
%% appropriate command.
%% Many elements are not rendered, but should be provided for metadata
%% extraction tools.

\author{Florentin Guth}
\affiliation{
  \institution{ETH Zurich, Switzerland}
}
\affiliation{
  \institution{{\'E}cole Normale Sup{\'e}rieure, France}
}
\email{florentin.guth@ens.fr}

\author{Valentin W{\"u}stholz}
\affiliation{
  \institution{ETH Zurich, Switzerland}
}
\email{wuestholz@gmail.com}

\author{Maria Christakis}
\affiliation{
  \institution{MPI-SWS, Germany}
}
\email{maria@mpi-sws.org}

\author{Peter M{\"u}ller}
\affiliation{
  \institution{ETH Zurich, Switzerland}
}
\email{peter.mueller@inf.ethz.ch}

\begin{abstract}
Smart contracts are programs that manage digital assets
according to a certain protocol, expressing for instance
the rules of an auction. Understanding the possible behaviors
of a smart contract is difficult, which complicates development,
auditing, and the post-mortem analysis of attacks.

This paper presents the first specification mining technique for smart
contracts. Our technique extracts the possible behaviors of smart
contracts from contract executions recorded on a blockchain and
expresses them as finite automata. A novel dependency analysis allows
us to separate independent interactions with a contract. Our technique
tunes the abstractions for the automata construction automatically
based on configurable metrics, for instance, to maximize readability
or precision. We implemented our technique for the Ethereum blockchain
and evaluated its usability on several real-world contracts.
\end{abstract}

%% ADD FOR CAMERA-READY
%% %% 2012 ACM Computing Classification System (CSS) concepts
%% %% Generate at 'http://dl.acm.org/ccs/ccs.cfm'.
%% \begin{CCSXML}
%% <ccs2012>
%% <concept>
%% <concept_id>10011007.10011006.10011008</concept_id>
%% <concept_desc>Software and its engineering~General programming languages</concept_desc>
%% <concept_significance>500</concept_significance>
%% </concept>
%% <concept>
%% <concept_id>10003456.10003457.10003521.10003525</concept_id>
%% <concept_desc>Social and professional topics~History of programming languages</concept_desc>
%% <concept_significance>300</concept_significance>
%% </concept>
%% </ccs2012>
%% \end{CCSXML}

%% \ccsdesc[500]{Software and its engineering~General programming languages}
%% \ccsdesc[300]{Social and professional topics~History of programming languages}
%% %% End of generated code

%% Keywords
%% comma separated list
%% \keywords{keyword1, keyword2, keyword3}  %% \keywords are mandatory in final camera-ready submission

%% \maketitle
%% Note: \maketitle command must come after title commands, author
%% commands, abstract environment, Computing Classification System
%% environment and commands, and keywords command.
\maketitle

%% ----------------------------------------------------------------
\section{Introduction}
\label{sect:Intro}
%% ----------------------------------------------------------------

Smart contracts are programs that store and automatically move digital
assets according to specified rules. While this idea was proposed over
20 years ago~\cite{Szabo1996}, it only gained traction when it was
combined with blockchain technology to store an immutable record of
all contract executions. Smart contracts have a wide range of
applications, including fund raising, securities trading and
settlement, supply-chain management, and electricity sourcing.

Despite their conceptual simplicity, understanding how to correctly
interact with a smart contract is often challenging, not only for
developers but also auditors. Specifically, legal invocations of
contract operations typically need to satisfy implicit temporal
ordering constraints, such as bidding only before an auction has
ended. As an additional complication, contract executions may have
subtle interactions (similar to data races) with each other when
accessing the same state, especially since the smart-contract
execution model provides no scheduling guarantees.

Specification mining~\cite{RobillardBodden2013} has been shown to help
software engineers understand behaviors of complex systems. In this paper, we
present the first application of specification mining to smart
contracts. Our technique can be used to understand a contract of
interest and its interactions with users and other smart contracts. We
mine these specifications from contract executions recorded on a
blockchain and express them as finite automata, describing the
observed sequences of interactions with the contract.

The automata produced by our technique not only characterize the
protocol (API) for using a contract, but also shed light on temporal
dependencies between different contract invocations that access the
same state. This is useful for understanding the functionality of a
smart contract, for debugging it, or even for a post-mortem analysis
of an attack. For instance, a bug in the Parity
wallet~\cite{ParityBug1} allowed a function for setting the wallet
owner to be called by anyone, even after the wallet was
constructed. As a result, attackers managed to claim existing wallets
by setting themselves as owners. In an automaton generated by our
technique, the contract invocation for updating the wallet owner would
appear as a transition that may be taken even after the wallet
construction, which would help developers understand the attack.

Our approach goes beyond existing specification mining techniques for
classical execution environments in two major ways. First, most
techniques target interactions of sequential clients with a data
structure~\cite{RobillardBodden2013}. In contrast, on a blockchain it
is common that multiple clients interleave their interactions with a
contract. To obtain precise specifications, our approach uses a novel
dependency analysis to separate independent interactions with a
contract and extract self-contained sequential traces from the
blockchain.

Second, existing mining techniques use fixed, built-in abstractions to
extract automata from sets of traces, e.g., by only considering method
signatures. To support a wide range of use cases, our approach
automatically adjusts its abstractions to maximize configurable
metrics. These metrics can, for instance, favor coarse abstractions to
obtain simple automata that are easy to grasp by a contract client, or
precise abstractions to facilitate accurate understanding by a human
auditor.

\begin{figure}[t!]
  \centering
  \scalebox{0.8}{
  \begin{tikzpicture}
    [->,>=Stealth,shorten >=1pt,auto,semithick]

    \node[state]                      (0)                      {0};
    \node[state,node distance=5cm]            (1) [right of=0] {1};
    \node[state,node distance=3cm,xshift=1cm] (2) [right of=1] {2};

    \path
    (0) edge node[outer sep=0.5cm,xshift=-0.5cm]{
      \begin{tabular}{|rl|}
        \hline
        Callee:    & $A$ \\
        Signature: & contract creation \\
        Input:     & \ldots \\
        \hline
      \end{tabular}
    } (1)
    (1) edge[bend left] node{
      \begin{tabular}{|rl|}
        \hline
        Caller:            & $*U$ \\
        Callee:            & $A$ \\
        Signature:         & \texttt{approve} \\
        \nth{1} parameter: & $*V$ \\
        \hline
      \end{tabular}
    } (2)
    (2) edge[bend left] node{
      \begin{tabular}{|rl|}
        \hline
        Caller:            & $V$ \\
        Callee:            & $A$ \\
        Signature:         & \texttt{transferFrom} \\
        \nth{1} parameter: & $U$ \\
        \nth{2} parameter: & $V$ \\
        \hline
      \end{tabular}
    } (1);
  \end{tikzpicture}}
  \vspace{-0.5em}
  \caption{An automaton generated by our technique.}
  \label{fig:Example-Automaton}
  \vspace{-2em}
\end{figure}
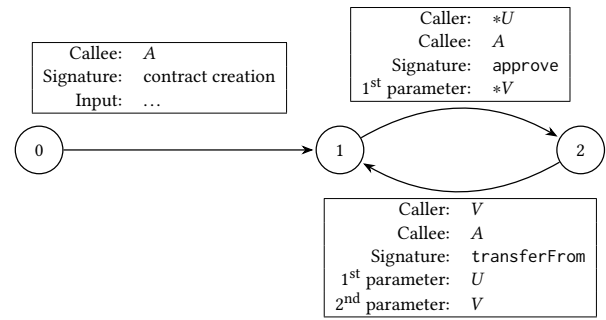

An example automaton that is generated by our technique is shown in
Fig.~\ref{fig:Example-Automaton}. This automaton represents part of
the functionality of the ERC20 token standard~\cite{ERC20}, an API
that allows clients to create and manage their own crypto-currency
using smart contracts. The edge labels of the automaton correspond to
contract invocations. Specifically, function \code{approve} of
contract instance $A$ is invoked by user $U$ to give user $V$
permission to withdraw assets from $U$'s account. Function
\code{transferFrom} of the same contract instance $A$ is invoked by
$V$ to withdraw assets from $U$'s account after $U$'s permission has
been granted. This functionality is depicted in the automaton of
Fig.~\ref{fig:Example-Automaton} through relational abstractions that,
for example, express that the first parameter of \code{transferFrom}
needs to match the caller of \code{approve}. The notation $*U$ and
$*V$ means that $U$ and $V$ are assigned fresh values. Although there
may be multiple concrete instances of $U$ and $V$, the automaton
expresses that \code{transferFrom} still expects the same instance of
$U$ and $V$ as in the previous invocation of \code{approve}, similarly
to named capture groups in regular expressions.

The automaton is readable because our dependency analysis extracts
self-contained interactions with the smart contract even though the
blockchain contains many (partly overlapping) interactions. The
abstractions applied here preserve relations between parameters of the
above contract invocations, but omit other information, such as the
value being approved and transferred or the return values of the
invocations.

Our work makes the following technical contributions:

\begin{enumerate}[1.]
\item We present the first application of specification mining to
  smart contracts.

\item We propose a novel dependency analysis that handles multiple
  interleaving interactions with a smart contract to extract
  self-contained sequential traces (\secref{sect:Histories}).

\item We present a novel automaton construction technique that
  automatically tunes its abstractions to optimize configurable
  metrics (\secref{sect:Automaton}).

\item We implemented our approach for the Ethereum blockchain and
  demonstrated its usefulness on several real-world contracts
  (\secref{sect:Experiments}).
\end{enumerate}

%% ----------------------------------------------------------------
\section{Background on Smart Contracts}
%% ----------------------------------------------------------------

A
\emph{blockchain}~\cite{BlockchainBlueprint,BlockchainTechnology,BlockchainRevolution}
is a decentralized consensus mechanism that was introduced by
Bitcoin~\cite{BitcoinMaster,BitcoinUnderstanding,BitcoinIntro}. More
specifically, a blockchain is a Byzantine fault-tolerant distributed
database that is replicated across a peer-to-peer network of nodes and
stores an ever-growing sequence of \emph{blocks}, each uniquely
identified by an increasing \emph{block number}. A subset of the
network nodes act as \emph{miners}; they collect, in a block, a
sequence of \emph{transactions}, which are broadcast to the network
but have not yet been stored in the blockchain. Transactions are
created by user nodes of the network, for example, to transfer
crypto-assets between different parties, and are communicated to the
network. To earn the right to append a block to the blockchain, a
miner needs to solve a mathematical challenge, in which case it
permanently stores the block in the blockchain with a link to the
previous block.

In the last few years, there have emerged several general-use,
blockchain-based, distributed-computing
platforms~\cite{BartolettiPompianu2017}, the most popular of which is
\emph{Ethereum}~\cite{EthereumWhitePaper}. Ethereum is open
source~\cite{Ethereum}, and its underlying crypto-currency is called
\emph{ether}. A key feature of Ethereum is its support for
\emph{contract accounts} in addition to \emph{user accounts}. Like
normal bank accounts, both contract and user accounts store a balance
in ether and are owned by a user. Both types of accounts publicly
reside on the Ethereum blockchain. A contract account, however, is not
directly managed by users, but rather through code that is associated
with it. Such code expresses contractual agreements between users, for
instance, to implement and enforce an auction protocol. A contract
account can also store persistent state (in a dictionary) that the
code may access, for instance, to store auction bids. To better
understand the process, imagine that a user issues a transaction with
the auction contract account to place a bid. When this transaction is
collected by a miner, the code of the contract account is
automatically executed and the bid is recorded in the state of the
account.

Contract accounts with their associated code and state are called
\emph{smart contracts}. The code is written in a Turing-complete
bytecode, which is executed on the Ethereum Virtual Machine
(EVM)~\cite{EthereumYellowPaper}. Of course, programmers do not
typically write EVM code. They can instead program in a variety of
high-level languages, such as Solidity, Serpent, or Vyper, which
compile to EVM bytecode.

%% ----------------------------------------------------------------
\section{Guided Tour}
\label{sect:Tour}
%% ----------------------------------------------------------------

In this section, we illustrate the workflow and architecture of our
specification mining approach for smart contracts. Through a running
example, we discuss the motivation behind the approach and the stages
of our technique.

\paragraph{Example.}
Fig.~\ref{fig:Example} shows a smart contract, written in Ethereum's
Solidity, that implements multiple, concurrent rock-paper-scissors
games. The contract provides a public API consisting of functions
\code{StartGame}, \code{Bet}, and \code{Claim}.

Function \code{StartGame} initializes a game
(line~\ref{line:gameInit}), specifying a period of 4 blocks (via
duration \code{d}), during which players may place their bets, and
returns the identifier of the new game (line~\ref{line:gameId}).
Function \code{Bet} requires the player to specify a game identifier
\code{gid}, their position \code{p} in the game (when \code{p} is 0,
the player is requesting to be \code{pA}, i.e., player A, and when
\code{p} is 1, they want to be \code{pB}), and their hand \code{h} (a
hand of 1 is rock, 2 is paper, and 3 is scissors). Additionally,
\code{Bet} is a payable function that expects players to pay the bet
amount (line~\ref{line:money}) within the first 4 blocks of the game
(line~\ref{line:block}). Function \code{Claim} allows players to claim
their winnings for a period of 4 blocks following the betting period
(lines~\ref{line:claimStart}--\ref{line:claimEnd}), and transfers money
to the winner (lines~\ref{line:pBRet}, \ref{line:pARet},
\ref{line:noWin}, \ref{line:pBWins},
and~\ref{line:pAWins}). Line~\ref{line:bugFix} disallows players from
claiming their winnings multiple times by manipulating the start of
claims such that, after the first claim, it appears that the claiming
period is already over.

\begin{figure}[t]
\begin{lstlisting}[language=Solidity]
contract RockPaperScissors {
  struct Game {
    address pA; address pB; // players
    uint hA; uint hB;       // hands
    uint cS;                // claim start
  }
  uint constant d = 4; uint constant amnt = 42;
  uint gC = 0;              // game count
  mapping(uint => Game) public games;

  function StartGame() public returns (uint) {
    gC++;
    var g = Game(0, 0, 0, 0, block.number + d); ¤\label{line:gameInit}¤
    games[gC] = g;
    return gC; ¤\label{line:gameId}¤
  }

  function Bet(uint gid, uint p, uint h) public payable {
    require(0 < h && h < 4 && p < 2);
    require(msg.value == amnt); ¤\label{line:money}¤
    var g = games[gid];
    require(0 < g.cS && block.number < g.cS); ¤\label{line:block}¤
    if (g.hA == 0 && p == 0) {
      g.pA = msg.sender; g.hA = h;
    } else if (g.hB == 0 && p == 1) {
      g.pB = msg.sender; g.hB = h;
    } else { require(false); }
  }

  function Claim(uint gid) public {
    var g = games[gid];
    require(0 < g.cS && g.cS <= block.number); ¤\label{line:claimStart}¤
    require(block.number < g.cS + d); ¤\label{line:claimEnd}¤
    g.cS = 0; // disallows multiple claims ¤\label{line:bugFix}¤
    if (g.hA == 0 && g.hB != 0) { // no player A
      g.pB.transfer(amnt); return; ¤\label{line:pBRet}¤
    }
    if (g.hB == 0 && g.hA != 0) { // no player B
      g.pA.transfer(amnt); return; ¤\label{line:pARet}¤
    }
    if (g.hA == g.hB) {           // draw
      g.pA.transfer(amnt); g.pB.transfer(amnt); ¤\label{line:noWin}¤
      return;
    }
    if (winningHand(g.hA, g.hB) == g.hB) {
      g.pB.transfer(2 * amnt); return; ¤\label{line:pBWins}¤
    }
    if (winningHand(g.hA, g.hB) == g.hA) {
      g.pA.transfer(2 * amnt); return; ¤\label{line:pAWins}¤
    }
  }
}
\end{lstlisting}
\vspace{-0.5em}
\caption{Running example written in Solidity.}
\vspace{-0.5em}
\label{fig:Example}
\end{figure}

\begin{figure}[t]
\centering
\scalebox{0.82}{
  \begin{tikzpicture}[align=center, node distance=1.6cm]
    \node[draw=none] (ST) at (0,0) {Seed transaction $T_S$};
    \node[draw, rounded corners=3, fill=black!5, draw=black!50, below of=ST, yshift=0.5cm] (HM) {History Mining};
    \node[draw, rounded corners=3, fill=black!10, draw=black!50, below of=HM, yshift=-3cm, dashed] (AO) {
    \begin{minipage}[t][5.2cm]{9cm}
    Automaton construction with abstraction tuning
    \end{minipage}};
    \node[draw, rounded corners=3, fill=black!20, draw=black!50, below of=AO, yshift=2.5cm, dashed] (AC) {
    \begin{minipage}[t][2cm]{7cm}
    Construction
    \end{minipage}};
    \node[draw, rounded corners=3, fill=black!5, draw=black!50, below of=AC, yshift=1.5cm] (B) {Construct\\automaton};
    \node[draw, rounded corners=3, fill=black!5, draw=black!50, left of=B, xshift=-0.5cm] (EA) {Abstract\\histories};
    \node[draw, rounded corners=3, fill=black!5, draw=black!50, right of=B, xshift=0.5cm] (AM) {Apply\\moves};
    \node[draw, rounded corners=3, fill=black!5, draw=black!50, below of=AC, yshift=-1.3cm] (S) {Sampling};
    \node[draw, rounded corners=3, fill=white, draw=white, yshift=-2.8cm, below of=AO] (user) {\scalebox{0.3}{\tikz{\draw[fill=black!50, draw=black!2] (-1.6cm,-1.6cm) rectangle (1.6cm,1.6cm); \draw[fill=black!2, draw=black!2] (0cm,0cm) ellipse (0.75cm and 1.05cm);\draw[fill=black!2, draw=black!2] (-1.4cm,-1.6cm) parabola bend (0cm,-1.2cm) (1.4cm,-1.6cm); \draw[fill=black!2, draw=black!2] (-0.3cm,-1.4cm) rectangle (0.3cm,-0.8cm);}}\\User};
    \draw[->, shorten >=1pt] (EA) -- (B);
    \draw[->, shorten >=1pt] (B) -- (AM);
    \draw[->, shorten >=1pt] (ST) -- (HM);
    \draw[->, shorten >=1pt, out=45, in=205, looseness=0.4] (AC.south west) to (EA.west);
    \draw[->, shorten >=1pt, out=45, in=205, looseness=0.4] (AC.south west) to (AM.south);
    \draw[->, shorten >=1pt, out=-25, in=110, looseness=0.4] (AM.east) to (AC.south east);
    \draw[thick, ->, shorten >=1pt, out=-125, in=25, looseness=1.4] (AC.south east) to node[midway, anchor=center, fill=black!10] {candidate automaton $A_{cand}$} (S.north east);
    \draw[thick, ->, shorten >=1pt, out=155, in=-45, looseness=1.4] (S.north west) to
    node[midway, anchor=center, fill=black!10] {new candidate\\ recipe $R_{cand}$} (AC.south west);
    \draw[thick, ->, shorten >=1pt] (HM) -- (AO) node[midway, anchor=center, fill=white] {histories $H_0$, \ldots, $H_n$};
    \draw[thick, ->, shorten >=1pt, out=-125, in=25, looseness=0.4] ([xshift=3.5cm]AO.south) to node[midway, anchor=center, fill=white] {automaton $A$ from final recipe $R$} (user.north east);
    \draw[thick, ->, shorten >=1pt, out=155, in=-45, looseness=0.4] (user.north west) to node[midway, anchor=center, fill=white] {user configuration} ([xshift=-3.5cm]AO.south);
  \end{tikzpicture}
}
\vspace{-0.5em}
\caption{Overview of the workflow and tool architecture.}
\label{fig:Architecture}
\vspace{0em}
\end{figure}
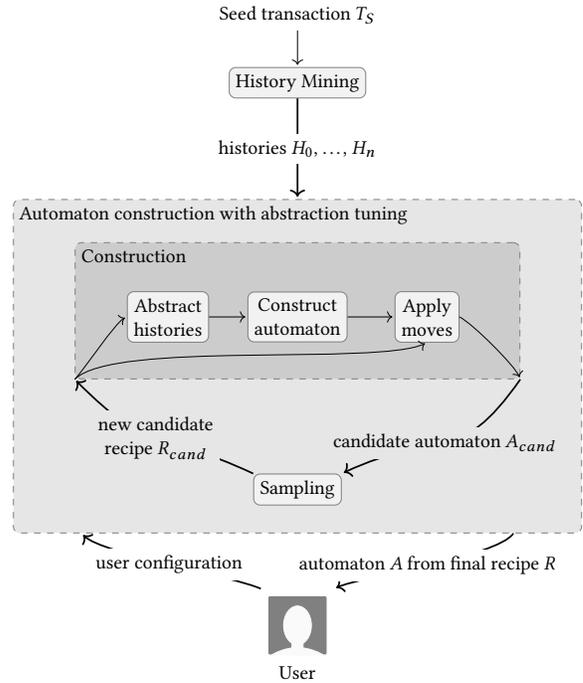

\paragraph{Workflow.}
Simple contracts such as our running example can be understood by
reading the source code. However, smart contracts are typically much
more complicated and their source code is often not available.  To
facilitate the understanding of even complex contracts---and thereby
their development, use, and auditing---we mine specifications that
characterize how contracts interact with users and other
contracts. For a given smart contract (the so-called \emph{target
  contract}), our technique determines these interactions by mining
actual executions that are recorded on the blockchain (either the
actual Ethereum blockchain or in a testing environment), and describes
them through a finite automaton.  As shown in
Fig.~\ref{fig:Architecture}, this process consists of two main steps,
which we explain next.

\paragraph{Mining histories.}
Our technique captures all interactions with the target contract
starting from a given \emph{seed transaction}, which contains an
invocation of the contract. In practice, the seed transaction
typically instantiates the target contract---similarly to how objects
are instantiated in object-oriented programs---such that all
interactions with the contract are captured. Our technique locates the
seed transaction on the blockchain and collects all subsequent
transactions. It then determines temporal dependencies among these
transactions, resulting in a directed \emph{dependency graph}. For
example, a transaction $T_2$ temporally depends on a transaction $T_1$
if $T_1$ occurs before $T_2$ on the blockchain and the execution of
$T_1$ affects the behavior of $T_2$, say, by writing a variable that
is read by $T_2$. Transactions that (transitively) depend on the seed
transaction should be captured in the generated automaton. All other
transactions are unrelated to understanding the target contract.
Tab.~\ref{tab:Transactions} shows a list of relevant transactions for
the smart contract from Fig.~\ref{fig:Example}, starting with the
instantiation of the contract. As an example, observe that $T_2$ from
Tab.~\ref{tab:Transactions} affects the behavior of $T_5$
by writing to location \code{games[g1]} that is read by $T_5$.

It is common that several clients interact simultaneously with an
instance of a smart contract. In fact, the transactions in
Tab.~\ref{tab:Transactions} come from four overlapping
rock-paper-scissors games. In order to obtain precise specifications,
it is important to separate these games into self-contained traces;
otherwise, transactions $T_4$ to $T_6$ would, for instance, suggest
that it is possible to call \code{Bet} three times in a row and modify
the hand of a player after it has been set.

To obtain this separation, we use the dependency graph to cluster
the relevant transactions into self-contained \emph{sessions}. A session
is the longest sequence of transactions such that the transactions in
the session depend only on other (earlier) transactions in the same
session and on the seed transaction. In our example, we identify a
session for each of the four rock-paper-scissors games in
Tab.~\ref{tab:Transactions}.

The dependency graph and sessions are defined on the granularity
of transactions since these are atomic operations on the
blockchain. Each transaction may include several contract
invocations. In order to obtain a precise automaton that also shows
the contract interactions within a transaction, we decompose each
transaction into its individual invocations. Performing this
decomposition on a session yields a sequence of invocations, called
\emph{history}. In Tab.~\ref{tab:Transactions}, each transaction
contains a single contract invocation, so the transformation of
sessions into histories is straightforward.

\paragraph{Constructing and optimizing the automaton.}
Once we have extracted a set of histories from the blockchain, we
represent them as a finite automaton. Histories here are sequences of
contract invocations, but our approach for constructing the automaton
is more general; it works for sequences of arbitrary \emph{events}
such as contract invocations, internal function calls, or statement
executions.

Each history can be represented trivially by an acyclic automaton that
represents a sequence of $n$ events by a chain of $n+1$ states. In
order to obtain more concise specifications, it is necessary to apply
abstractions, which may merge different events and, thereby,
facilitate the construction of smaller, cyclic automata.

Choosing suitable abstractions for the automaton construction is
difficult for two reasons. First, it depends on the intended purpose
of the generated automaton. A user of the target contract
might prefer a simple automaton that is easy to read and, thus, favor
coarse abstractions. On the other hand, an auditor might require
an increased precision when examining more
subtle contract interactions. Second, the space of possible
abstractions is huge. For instance, even for each argument of an event
(such as parameter and result values), there are numerous possible
abstractions that strike different cost-benefit ratios.

\begin{table}[t!]
\centering
\caption{A sequence of transactions with the \code{RockPaperScissors}
  contract (of Fig.~\ref{fig:Example}) starting from the
  contract-creation transaction.}
\vspace{-1em}
%%\scalebox{0.8}{
\begin{tabular}{c|c|c}
\textsc{Transaction} & \textsc{Contract}   & \textsc{Block}\\
\textsc{Identifier}  & \textsc{Invocation} & \textsc{Number}\\
\hline
 1 & contract creation       &  7\\
 2 & \code{g1 = StartGame()} &  9\\
 3 & \code{g2 = StartGame()} & 10\\
 4 & \code{Bet(g2, 1, 3)}    & 10\\
 5 & \code{Bet(g1, 0, 1)}    & 11\\
 6 & \code{Bet(g1, 1, 2)}    & 12\\
 7 & \code{Claim(g1)}        & 13\\
 8 & \code{g3 = StartGame()} & 14\\
 9 & \code{Claim(g2)}        & 14\\
10 & \code{Bet(g3, 0, 3)}    & 15\\
11 & \code{Bet(g3, 1, 1)}    & 15\\
12 & \code{g4 = StartGame()} & 16\\
13 & \code{Bet(g4, 1, 1)}    & 17\\
14 & \code{Bet(g4, 0, 1)}    & 17\\
15 & \code{Claim(g4)}        & 20\\
16 & \code{Claim(g4)}        & 20\\
\end{tabular}
%%}
\label{tab:Transactions}
\vspace{0em}
\end{table}

Instead of using fixed abstractions, we address this challenge by
optimizing the automaton according to a user-defined
configuration. This configuration could, for instance, favor
readability (e.g., by penalizing large automata) or precision (e.g.,
by penalizing large information loss). Our technique iteratively
adjusts the applied abstractions to obtain a useful, or even optimal,
result. This process is depicted by the outer gray box in
Fig.~\ref{fig:Architecture} and explained below.

\begin{figure}[t]
  \includegraphics[height=36em]{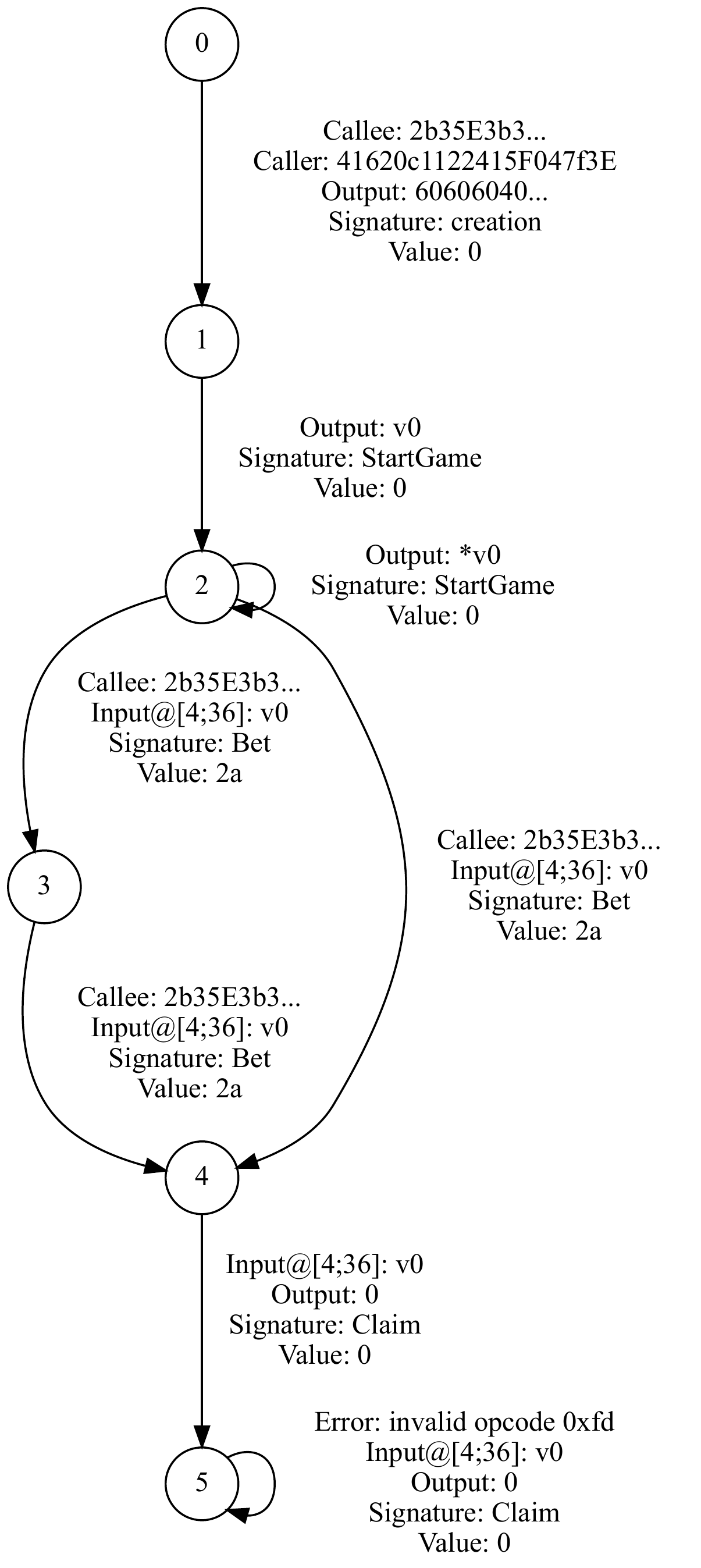}
  \vspace{-0em}
  \caption{The output automaton for the \code{RockPaperScissors}
    contract (of Fig.~\ref{fig:Example}), based on the
    transactions of Tab.~\ref{tab:Transactions}.}
  \label{fig:tunedAutomaton}
  \vspace{-0em}
\end{figure}

The automaton construction proceeds in three steps (see inner gray box
in Fig.~\ref{fig:Architecture}). First, starting from the set of mined
histories, we apply \emph{event abstractions} to each event in the
histories, resulting in a set of \emph{abstract histories}.  Event
abstractions omit details of concrete events, for instance, they might
ignore certain result values or abstract numerical parameters using
standard abstract domains. Second, the resulting set of abstract
histories is represented by a finite automaton. Third, this automaton
is further simplified by applying \emph{automaton moves}, which
express local, automaton-specific abstractions such as merging two
states.

To optimize the resulting automaton, we apply this construction
process iteratively, as follows. The event abstractions and automaton
moves are described by a so-called \emph{recipe}; the initial recipe
uses the identity event abstraction and contains no automata moves. In
each iteration, we construct a candidate automaton $A_{\mathit{cand}}$
according to the current candidate recipe.  The subsequent sampling
step then computes a cost for the candidate automaton
$A_{\mathit{cand}}$ based on a user-defined configuration and
remembers the best automaton constructed so far. In search for an even
better automaton, the sampling applies a random variation to the
recipe and proceeds using this new candidate recipe. This process
iterates until it reaches a user-defined exploration bound, and
returns the best automaton so far. The user may then inspect the
automaton and, if necessary, restart the process with an adjusted
configuration to obtain a more concise or a more detailed result.

For the running example, our tool chain produces the automaton of
Fig.~\ref{fig:tunedAutomaton} using a configuration that strikes a
good balance between readability and precision. This automaton is
constructed based on the transactions of Tab.~\ref{tab:Transactions}; the
events appear as edge labels.

The automaton shows that multiple games may be
started concurrently. The first \code{StartGame} event returns a
variable $v_0$ that represents the game identifier. The notation
$*v_0$ of the subsequent \code{StartGame} events expresses that each
new game has a different identifier. Following the \code{StartGame}
events, we have one or two consecutive invocations of \code{Bet} for a
game $v_0$, along the right and left paths of the automaton,
respectively. We then find a single successful \code{Claim} event for
$v_0$.
Here, the automaton shows that it is possible to invoke \code{Claim}
after only one \code{Bet}; this functionality is needed to refund the
bet amount if no second player participates in the game and emerges
because of \code{g2} of Tab.~\ref{tab:Transactions}.
Observe that any further invocations of \code{Claim} for the same game
$v_0$ result in an error, which is denoted by a self-loop in the
automaton.

The automaton abstracts over the concrete result of \code{StartGame} and the first
parameters of \code{Bet} and \code{Claim}. It also abstracts away the second and third
parameters of \code{Bet}, while not abstracting the transferred amounts of Ether (e.g.,
value 42 (0x2a) for invocations of \code{Bet}).
A characteristic of the automata that our technique generates is that
each state is an accepting state, expressing that an interaction with
the target contract may terminate at any state.  For example, it is
not necessary that a \code{StartGame} event is followed by an
invocation of \code{Bet} and \code{Claim}.
Our construction ensures that the resulting automaton
over-approximates the histories extracted from the blockchain. That
is, each of these histories is accepted by a run of the automaton.

In the following sections, we describe the main components of our
architecture in more detail and explain precisely how we generate the
automaton of Fig.~\ref{fig:tunedAutomaton}.

%% ----------------------------------------------------------------
\section{Mining Histories}
\label{sect:Histories}
%% ----------------------------------------------------------------

The process of extracting histories from the blockchain starts by
locating the given seed transaction $T_S$ on the blockchain and
collecting all subsequent transactions. These transactions are then
processed in four major steps.  First, to analyze the dependencies
among these transactions, we determine the read and write effects for
each of them (Sect.~\ref{subsect:Effects}). Second, we use these
effects to build a dependency graph, which describes temporal
dependencies between transactions that access the same state
(Sect.~\ref{subsect:Graph}). Third, we remove any transactions that
are unrelated to $T_S$ and, consequently, to the target contract
(Sect.~\ref{subsect:Filtering}). Fourth, we identify independent
sessions and then decompose the transactions of each session to
generate the histories of events that should be represented in the
constructed automaton (Sect.~\ref{subsect:Histories}).

\subsection{Collecting transaction effects}
\label{subsect:Effects}

The automata we produce summarize those transactions on the blockchain
that are related to a given seed transaction; other transactions are
irrelevant and should be omitted to obtain concise specifications. Two
transactions are related if they access the same memory
locations. Consequently, our analysis starts by collecting the read
and write effects of the seed and all subsequent transactions. For
this purpose, we execute each transaction on a local copy of the
blockchain and use hooks into the virtual machine to record all memory
accesses. This process obtains precise results, but is time consuming.
However, it needs to be performed only once for each transaction on
the blockchain; the results can be cached and extended incrementally
as the blockchain grows. Moreover, it is often useful to mine only a
small portion of the blockchain, for instance, when an auction lasts
at most one week.

In general a location is any persistent state that a transaction may
access. However, many transactions access the balance of
accounts. These accesses introduce relationships between transactions
that are otherwise unrelated and should be ignored in the mined
specifications.  For instance, they relate a game a user plays with an
investment they make.  To avoid such spurious connections, we exclude
account balances from the locations we consider in the following.
Technically, this means that we (slightly) under-approximate the
effects of each transaction.

\begin{definition}[Location]
A \emph{location} $l \in \mathcal{L}$ is any persistent state that the
execution of a transaction may access (i.e., read or write) and that
is not the balance of an account.
\end{definition}

A location may, for instance, be an index to the persistent dictionary
of a smart contract or an attribute of the block in which a
transaction is contained, e.g., the block number or timestamp.  For
instance, in the running example, \code{g.pA} and \code{block.number}
are locations accessed by transaction $T_5$ (from
Tab.~\ref{tab:Transactions}).
Using the above definition, we can now define the effects of a
transaction.

\begin{definition}[Write effect]
  The \emph{write effect} $w(T) \subseteq \mathcal{L}$ of a
  transaction $T$ is the set of locations that are written to
  during execution of $T$.
\end{definition}

For instance, the write effect $w(T_2)$ contains locations \code{gC},
\code{g.pA}, \code{g.pB}, \code{g.hA}, \code{g.hB}, \code{g.cS}, and
\code{games[gC]}.

\begin{definition}[Read effect]
\label{def:read-effect}
The \emph{read effect} $r(T) \subseteq \mathcal{L}$ of a transaction
$T$ is the set of locations that are read from during execution of $T$,
  \emph{without having previously been written to by $T$}.
\end{definition}

For instance, the read effect $r(T_6)$ contains \code{games[gid]},
\code{g.cS}, \code{block.number}, \code{g.hA}, and \code{g.hB}. Since
we will use effects to determine dependencies between different
transactions, Def.~\ref{def:read-effect} ignores reads from a location
that is previously written to by the same transaction. Consequently,
if $l \in w(T) \cap r(T)$, then a read from $l$ occurs before a write
to $l$ during execution of $T$, e.g., $\text{\code{g.hB}} \in w(T_6)
\cap r(T_6)$.

A transaction may read attributes of the block in which it is
contained (such as the block number). These attributes are updated
automatically when a block is appended to the blockchain.  In order to
reflect these implicit write operations, our technique inserts a
\emph{ghost transaction} right before the first transaction of every
block. The write effect of this ghost transaction contains all block
attributes. For instance, the ghost transaction for block~11, denoted
$B_{11}$, writes to location \code{block.number}, which is then read
by $T_5$.

\subsection{Building the dependency graph}
\label{subsect:Graph}

We use read and write effects to compute two kinds of dependencies
between transactions, called strong and weak dependencies. Strong
dependencies are used to determine the relevant transactions, that is,
the transactions that are related to the seed transaction and,
therefore, need to be reflected in the constructed automaton.  Weak
dependencies help define the order in which the relevant transactions
may occur.

\begin{definition}[Strong dependency] \label{def:StrongDataEdge}
A transaction $T_j$ \emph{strongly depends} on transaction $T_i$,
denoted by $T_i \rightarrow T_j$, if and only if (1)~$T_i$ occurs
before $T_j$ on the blockchain and (2)~$T_i$ writes to a location that
$T_j$ reads, and no transaction between $T_i$ and $T_j$ writes to that
location: $(w(T_i) \cap r(T_j)) \setminus B \neq \emptyset$, where $B
= \bigcup_{k = i+1, \ldots, j-1} w(T_k)$.
\end{definition}

Intuitively, a strong dependency reflects that the execution of $T_i$
affects the execution of $T_j$. Consequently, it indicates that $T_i$
should be included in the constructed automaton if $T_j$ is (we will
explain the details in the next subsection). Moreover, if they are
included, the automaton must reflect that $T_i$ occurs before $T_j$.

A weak dependency alone does not determine which transactions to
include in the automaton, but expresses an ordering constraint that
needs to be reflected; changing the order of two weakly dependent
transactions may influence the result of the execution.

\begin{definition}[Weak dependency] \label{def:WeakDataEdge}
A transaction $T_j$ \emph{weakly depends} on transaction $T_i$,
denoted by $T_i \dashrightarrow T_j$, if and only if (1)~$T_i$ occurs
before $T_j$ on the blockchain and (2)~$T_i$ writes to or reads from a
location that $T_j$ writes, and no transaction between $T_i$ and $T_j$
writes to that location: $(w(T_i) \cap w(T_j)) \setminus B \neq
\emptyset$ or $(r(T_i) \cap w(T_j)) \setminus B \neq \emptyset$, where
$B$ is defined as above.
\end{definition}

We refer to strong and weak dependencies as \emph{temporal
dependencies}; we use them to build a transaction dependency graph:

\begin{definition}[Dependency graph]
  A \emph{dependency graph} is a directed graph, in which each node
  represents a transaction and each edge a strong or weak dependency
  between two transactions.
\end{definition}

\begin{figure}
  \includegraphics[width=0.168\textwidth]{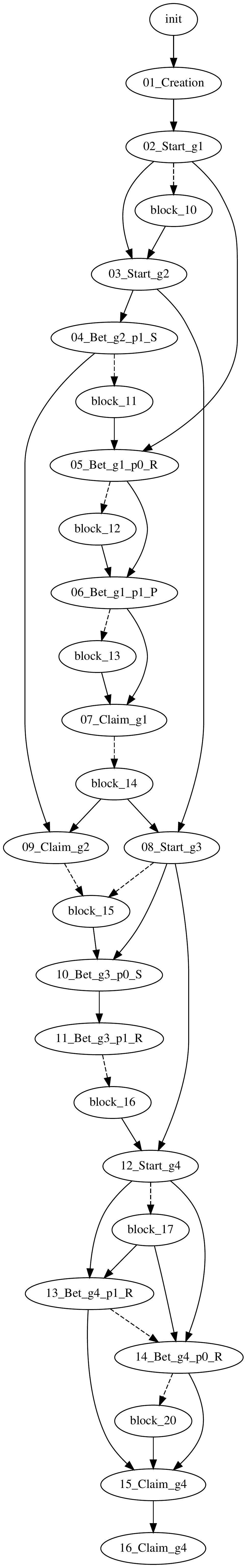}
  \caption{The dependency graph for the \code{RockPaperScissors}
    contract (of Fig.~\ref{fig:Example}), generated from the
    transactions of Tab.~\ref{tab:Transactions}. Each
  (non-ghost) node in the graph is annotated with the corresponding
  transaction identifier followed by the contract invocation with its
  arguments. The graph does not show edges between nodes when
  these are implied by transitivity of the dependencies.}
  \label{fig:DependencyGraph}
\end{figure}

The dependency graph generated by our tool chain for the running
example is shown in Fig.~\ref{fig:DependencyGraph}. Notice
that there is a strong dependency $T_5 \rightarrow T_6$ because function
\code{Bet} of $T_5$ writes to \code{g.hA}, which is read by
\code{Bet} of $T_6$. However, there is a weak dependency $T_{13} \dashrightarrow
T_{14}$ since \code{Bet} of $T_{13}$ reads from \code{g.hA}, which
is written to by \code{Bet} of $T_{14}$. There is also a weak dependency
$T_9 \dashrightarrow B_{15}$ (the ghost transaction that is inserted by
our technique right before the first transaction of block~15), which
indicates that function \code{Claim} of $T_9$
reads from location \code{block.number}, which is written to by
$B_{15}$.

%% To build a transaction dependency graph, our technique only considers
%% reads and writes to data locations, that is, excluding any reads and
%% writes to account balances. This means that the generated dependency
%% graph under-approximates the temporal dependencies between the seed
%% and all subsequent transactions (according to
%% Def.~\ref{def:TemporalDependency}). However, ignoring transaction
%% effects containing account balances is critical for mining precise and
%% informative specifications for a target smart contract, as we show
%% next with an example.

%% Assume that transaction $T_4$ (of Tab.~\ref{tab:Transactions}) was
%% issued by player John Doe, who after placing his bet, transferred some
%% assets to his friend Jane Roe. The transaction with Jane would
%% strongly depend on $T_4$ as John's balance is written to by $T_4$ and
%% read from by the transaction with Jane. We could then imagine other
%% asset transfers issued by Jane, which would also have a strong
%% (transitive) dependency on $T_4$ even though they are unrelated to the
%% rock-paper-scissors contract. As discussed at the beginning of this
%% section, our technique mines the histories that should be
%% characterized by the output automaton by traversing the dependency
%% graph. We, therefore, ignore accesses to account balances to avoid
%% creating dependencies with unrelated asset-transferring transactions
%% and prevent their events from showing up in the output automaton.

\subsection{Filtering irrelevant transactions}
\label{subsect:Filtering}

Intuitively, we consider a transaction to be relevant if: (a)~It is
(directly or transitively) affected by the execution of the seed
transaction, for example, a bid in an auction that was started by the
seed transaction. (b)~It occurs after the seed transaction and it
(directly or transitively) affects the execution of an already
relevant transaction according to case~(a), for instance, the
transaction that initializes the minimum price of an auctioned item
before a bid is placed in an auction that was started by the seed
transaction. In order to formalize this intuition, we first define
transitive dependencies.

\begin{definition}[Strong dependency path]
A \emph{strong dependency path} from transaction $T_i$ to $T_j$ is
denoted $T_i \rightarrow^* T_j$ and represents a non-empty path from
$T_i$ to $T_j$ consisting only of strong dependencies.
\end{definition}

\begin{definition}[Weak dependency path]
A \emph{weak dependency path} from transaction $T_i$ to $T_j$ is
denoted $T_i \; \dashrightarrow^* \; T_j$ and represents a non-empty
path from $T_i$ to $T_j$ consisting of any number of strong
dependencies and at least one weak dependency.
\end{definition}

We remove all irrelevant transactions from the dependency graph as
described by the following definition.

\begin{definition}[Filtered dependency graph] \label{def:Filtering}
  Given a dependency graph $G$, a \emph{filtered dependency graph} is
  generated from $G$ by removing any transaction $T$  that
  does not fulfill either of the following two conditions:
  \begin{enumerate}[a.]
  \item There exists a strong dependency path $T_S \rightarrow^* T$.
  \item There exists a weak dependency path $T_S \dashrightarrow^* T$, and
    there exists a transaction $T'$ such that there exist strong dependency
    paths $T \rightarrow^* T'$ and $T_S \rightarrow^* T'$.
  \end{enumerate}
\end{definition}

This definition closely matches the intuition given at the beginning
of this subsection. The requirement of a weak dependency path in
case~(b) is motivated by the fact that without at least a weak path,
transactions $T_S$ and $T$ would be unordered. If they were unordered,
$T$ might as well have occurred \emph{before} the seed
transaction. Therefore, considering $T$ to be a relevant transaction
would contradict the role of the seed transaction as the ``beginning
of time''. However, this requirement could in principle be dropped if
a more relaxed interpretation of the seed transaction was desired.

%% The filtering removes transactions with no
%% dependency on the seed because these transactions might as well have
%% been executed before $T_S$. Therefore, they cannot be essential for
%% describing the functionality of the target contract when starting from
%% $T_S$. Our filtering also removes transactions that have a
%% weak dependency on the seed but do not satisfy condition (b) from
%% above. However, this scenario does not occur for non-ghost
%% transactions when $T_S$ instantiates the target contract. Only
%% invocations of the target contract may access its persistent state,
%% but all such invocations have a strong dependency on the seed (because
%% they read the contract bytecode). Consequently, the transactions we
%% remove from the dependency graph typically reflect noise that would
%% make the construction of the output automaton less precise.

In the dependency graph of Fig.~\ref{fig:DependencyGraph}, we do not
filter out the ghost transaction $B_{11}$ because it satisfies case
(b) of the above definition: It has a weak dependency on the seed
$T_1$ and the strong dependency paths $B_{11} \rightarrow T_5$ and
$T_1 \rightarrow^* T_5$. If \code{Bet} was not required to be invoked
within the betting period (line~\ref{line:block} of
Fig.~\ref{fig:Example}), there would be no edge between $B_{11}$ and
$T_5$, in which case transaction $B_{11}$ would be filtered out.
As another example, imagine that we create a second instance of the
rock-paper-scissors contract. Even if the same players participated in
games of both contracts, the accessed locations would be disjoint
because we exclude account balances when determining read and write
effects. As a result, all transactions with the second instance of the
contract would not depend on the seed $T_1$ and would be removed.

\subsection{Generating histories of events}
\label{subsect:Histories}

In the final phase of the history mining, our technique traverses the
filtered dependency graph to cluster its transactions into
self-contained sessions. The transactions in these sessions are then
decomposed to obtain histories of events.  The events (that is,
contract invocations) are collected when our technique executes the
transactions on a local copy of the blockchain to compute their read
and write effects (see \secref{subsect:Effects}).

Sessions represent independent interactions with a contract, for
instance, different rock-paper-scissors games. Since strong
dependencies indicate that two transactions are related by influencing
each other's execution, strongly dependent transactions belong to the
same session:

\begin{definition}[Session]
  A \emph{session} is the longest sequence of transactions that have a
  strong dependency only on transactions in the same session.
\end{definition}

Intuitively, each transaction that is a sink in the graph with respect
to strong dependencies (called \emph{final transaction} below) can be
placed in a separate session because no other transaction strongly
depends on it. A session containing a final transaction $T_F$ contains
all transactions $T$ such that $T \rightarrow^* T_F$.

Sessions are sequences of transactions, that is, ordered.  The mined
specification should reflect all possible orderings of the
transactions in a session that are consistent with the executions
extracted from the blockchain. These orderings are reflected by the
strong and weak dependencies in the filtered dependency graph. We
capture them by introducing potentially multiple sessions per final
transaction: one for each topological ordering of the transactions.

For instance, there are four final transactions in the dependency
graph of Fig.~\ref{fig:DependencyGraph}, namely $T_7$, $T_9$,
$T_{11}$, and $T_{16}$, giving rise to the following four sessions.
Note that the sessions for different final transactions overlap, at
least in the seed transaction.
\[
\begin{array}{lll}
S_1 & := & [T_1, T_2, B_{11}, T_5, B_{12}, T_6, B_{13}, T_7]\\
S_2 & := & [T_1, T_2, B_{10}, T_3, T_4, B_{14}, T_9]\\
S_3 & := & [T_1, T_2, B_{10}, T_3, B_{14}, T_8, B_{15}, T_{10}, T_{11}]\\
S_4 & := & [T_1, T_2, B_{10}, T_3, B_{14}, T_8, B_{16}, T_{12}, B_{17}, T_{13}, T_{14}, B_{20}, T_{15}, T_{16}]\\
\end{array}
\]
Although weak dependencies are not considered when computing which
transactions should be included in a session, they are useful in
determining the ordering of transactions in sessions. For instance,
when computing session $S_1$, observe that both $B_{12}$ and $T_5$
have a strong dependency path to $T_6$, but only the weak dependency
from $T_5$ to $B_{12}$ reflects in which order these transactions
occur. Ignoring such weak dependencies would lead to more sessions,
some of which might correspond to invalid executions.

It might be tempting to restrict the order of transactions in a
session to exactly the order in which they occur on the blockchain.
However, this approach would lead to sessions that reflect the
accidental ordering of unrelated transactions and, thus, miss other
orderings that would have been possible and, therefore, should be
shown in the mined specification.

Even though transactions are the atomic units of execution on the
blockchain, it is more informative to express specifications of smart
contracts on the level of individual events, in particular, contract
invocations. For this purpose, we decompose the transactions in each
session into its constituent events to obtain a set of histories:

\begin{definition}[History]
  A \emph{history} is a sequence of events that occur during the execution
  of a session.
\end{definition}

Note that we perform the filtering, clustering, and ordering on the
level of transactions and only then decompose the transactions into
events. Since the events within one transaction are typically strongly
dependent, performing the processing on the level of individual events
would not increase the precision of our specifications, but lead to a
much higher computational effort.

For the running example, we generate four histories of events by
replacing each transaction in the above sessions with the
corresponding contract invocation (from Tab.~\ref{tab:Transactions}).
Note that a ghost transaction does not contain any contract
invocations.

%% ----------------------------------------------------------------
\section{Automaton Construction with Automatic Abstraction Tuning}
\label{sect:Automaton}
%% ----------------------------------------------------------------

Finding suitable abstractions to apply during the automaton
construction is difficult. As discussed earlier, not only is the space
of abstractions huge, but also the suitability of an abstraction
depends on the intended use of the automaton. To address these
challenges, our technique automatically tunes its abstractions to
maximize the user-defined configuration, like readability or
precision.

\paragraph{Algorithm.}
Alg.~\ref{alg:AutomatonOptimizer} presents the automaton construction
with abstraction tuning. As shown in Fig.~\ref{fig:Architecture}, it
takes as input a set of concrete histories and builds a candidate
automaton $A_{cand}$ from the input histories and a candidate recipe
$R_{cand}$.  Recall from Sect.~\ref{sect:Tour} that a recipe is a
sequence of event abstractions and automaton moves. Event abstractions
and automaton moves are applied as pre- and post-processing steps of
the actual automaton construction, respectively (see procedure
\textsc{ApplyAutomatonRecipe}).  The actual construction is
straightforward and results in a tree-shaped automaton that precisely
characterizes all (abstracted) input histories. Cycles are introduced
later, when automaton moves are applied.  The initial recipe consists
only of the identity event abstraction.

\begin{algorithm}[t]
  \caption{Automaton construction with abstraction tuning.}
  \label{alg:AutomatonOptimizer}
  \begin{algorithmic}[1]
%%    \small
%%    \fontsize{7}{15}
    \Procedure{ConstructAndTuneAutomaton}{$H_0$, \ldots, $H_n$}
      \Let{$R_{cand}$}{\Fcall{IdentityEventAbstraction}()}
      \Let{$C_{opt},C_{lst}$}{$\infty$}
      \Let{$R_{opt},R_{lst}$}{$R_{cand}$}
      \While{$\neg$\Fcall{BoundReached}()}
         \Let{$A_{cand}$}{\Fcall{ApplyAutomatonRecipe}($H_0$, \ldots, $H_n$, $R_{cand}$)}
        \Let{$C_{cand}$}{\Fcall{ComputeAutomatonCost}($A_{cand}$)}
        \If{$C_{cand} < C_{opt}$}
          \Let{$C_{opt}, R_{opt}$}{$C_{cand}, R_{cand}$}
        \EndIf
        \If{\Fcall{Accept}($C_{cand}, C_{lst}$)}
          \Let{$C_{lst},R_{lst}$}{$C_{cand}, R_{cand}$} \label{line:accept}
        \EndIf
        \Let{$R_{cand}$}{\Fcall{ModifyRecipe}($R_{lst}$)}
      \EndWhile
      \State \textbf{return} {\Fcall{ApplyAutomatonRecipe}($H_0$, \ldots, $H_n$, $R_{opt}$)}
    \EndProcedure
    \Procedure{ApplyAutomatonRecipe}{$H_0$, \ldots, $H_n$, $R$}
      \Let{$H_0'$, \ldots, $H_n'$}{\Fcall{AbstractHistories}($H_0$, \ldots, $H_n$, $R$)}
      \Let{$A_{tmp}$}{\Fcall{BuildAutomaton}($H_0'$, \ldots, $H_n'$)}
      \State \textbf{return} {\Fcall{ApplyAutomatonMoves}($A_{tmp}$, $R$)}
    \EndProcedure
  \end{algorithmic}
\end{algorithm}

Once the initial automaton has been constructed, our algorithm tries
to optimize it, according to a user-provided configuration. For this
purpose, it applies random variations to the recipe, hoping to achieve
a lower cost. A key insight of our algorithm is that these variations
are not applied to the best recipe seen so far, but to a recent
one. This approach allows the algorithm to explore recipes that might
temporarily increase the cost of the candidate automaton but
eventually lead the exploration away from a local minimum.

The algorithm keeps track of the current recipe ($R_{cand}$), the best
recipe so far ($R_{opt}$), and the recipe used for the next random
variation ($R_{lst}$), together with their associated costs.  After
computing a candidate automaton, the algorithm compares its cost to
the best recipe so far and updates the information if a new best
automaton has been found. It then decides which recipe to accept as a
basis for the next iteration. $R_{cand}$ is chosen if its cost is
lower than the cost for the previous $R_{lst}$; even if the cost is
higher, it is accepted with a certain probability that decreases
proportionally to how much the cost increases as well as to how much
time has elapsed.  This process allows the algorithm to steer away
from local minima, but also ensures that the exploration stabilizes
over time.
This iterative process, inspired by \emph{simulated
  annealing}~\cite{MetropolisRosenbluth1953,KirkpatrickGelatt1983},
is stopped when an exploration bound is reached, such as the
maximum number of iterations or a timeout. When stopped, the user
is presented with the best automaton so far.

\begin{figure}[t]
\centering
\scalebox{0.8}{
\begin{tikzpicture}
    \begin{axis}[
            width=0.6\textwidth,
            height=.25\textwidth,
            xtick={1,1000,2000,3000,4000,5000},
            ytick={0,10,20,30,40,50},
            ymin=0,ymax=55,
            xmin=1,xmax=5000,
            xlabel={\textbf{Number of sampling steps}},
            ylabel={\textbf{Cost}},
            axis y line*=none,
            axis x line*=bottom,
        ]
        \addplot[very thick,blue] table [x=Step, y=Cost, col sep=comma] {figures/cost-plot-small.csv};
    \end{axis}
\end{tikzpicture}}
\vspace{-1.5em}
\caption{The cost of the automaton of Fig.~\ref{fig:tunedAutomaton}
  for the first 5000 sampling steps.}
\label{fig:costPlot}
\vspace{-0em}
\end{figure}
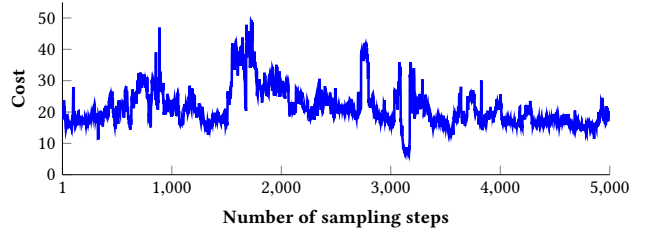

Fig.~\ref{fig:costPlot} plots the cost of the automaton of Fig.~\ref{fig:tunedAutomaton}
over the number of sampling steps. Observe that our algorithm avoids several local
minima by allowing the cost to increase up to $\sim$50 before dropping to
6.7. As the number of steps increases, the cost is allowed to increase less. After 5,000
steps, the cost is $\sim$20, which is higher than the best cost of 6.7 observed around
step 3,100. As discussed earlier, our algorithm returns the automaton with the best cost.

\paragraph{Cost metrics.}
The metrics used by our algorithm are essentially a linear combination
of readability, generality, and precision. In particular, it is
possible to penalize automata with many states and edges to favor
readability, to reward automata that describe more histories than
those directly observed on the blockchain to favor generality, and to
penalize automata that describe histories that were not observed on
the blockchain to favor precision.

\begin{figure}[t]
  \includegraphics[height=37em]{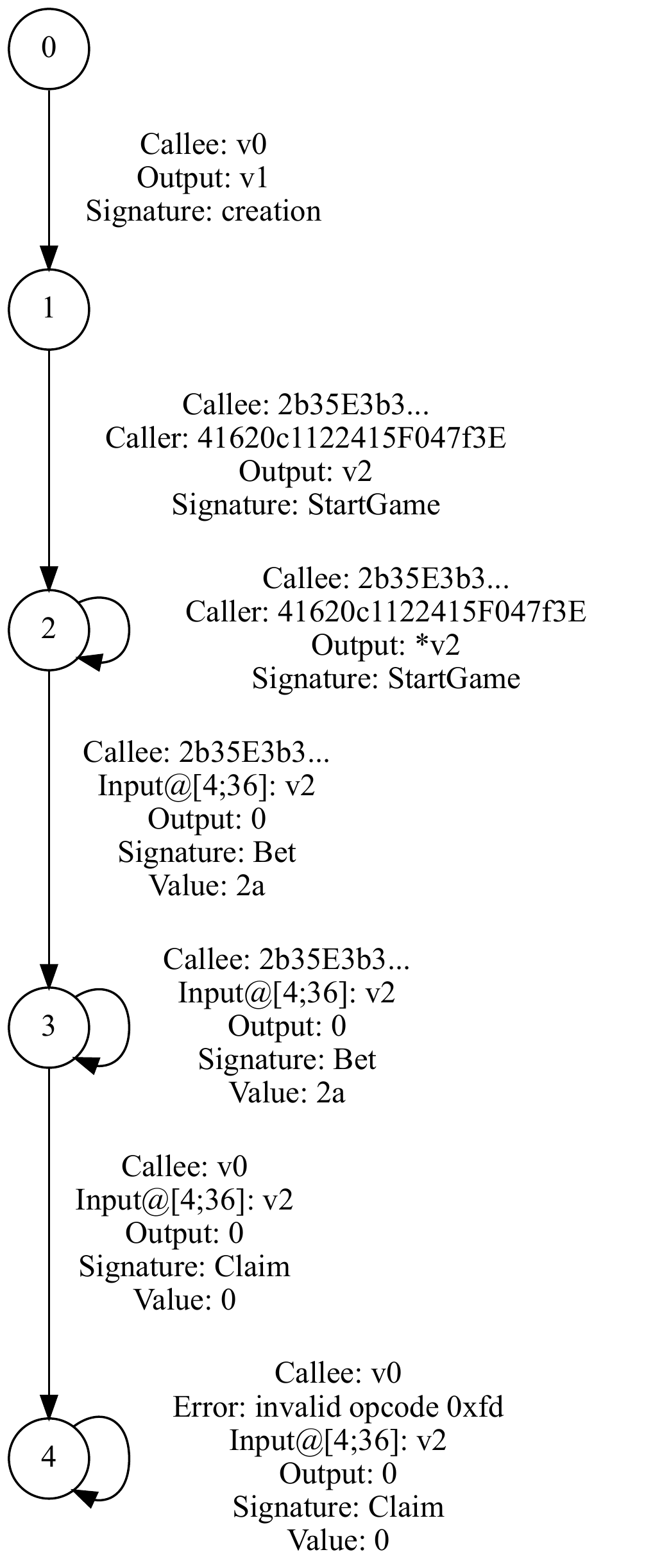}
  \vspace{0em}
  \caption{A general automaton for the contract of
    Fig.~\ref{fig:Example}.}
  \label{fig:generalAutomaton}
  \vspace{0em}
\end{figure}

Fig.~\ref{fig:generalAutomaton} shows an automaton generated by our
technique for the rock-paper-scissors contract when the user
configuration favors simplicity and generality over precision.  The
automaton still shows, for instance, that \code{Claim} events occur only
after at least one \code{Bet} event. However, it now allows one or
more bets instead of one or two, as in the more precise automaton from
Fig.~\ref{fig:tunedAutomaton}.

\paragraph{Event abstractions.}
Our technique represents events as records whose fields store
information about their occurrence at execution time, for instance
an \code{Output} field stores the return value of a contract
invocation, as in
Fig.~\ref{fig:tunedAutomaton}. In order to obtain concise
  automata, we allow our algorithm to abstract from the
  details using the following event abstractions. A recipe
  determines the abstraction per function signature and event field.
\begin{enumerate}
\item[--] \emph{Identity abstraction}: The value of the field is unchanged.
\item[--] \emph{Variable abstraction}: The value is abstracted by a
  variable.
\item[--] \emph{Top abstraction}: The value is abstracted away.
\end{enumerate}

To obtain the automaton from Fig.~\ref{fig:tunedAutomaton},
our technique applied, for instance, the variable abstraction
on the \code{Output} field of all \code{StartGame}
events. This means that the \code{Output} field of these events was
each assigned a fresh variable. These variables were then replaced by
a single variable $v_0$ using an automaton move, which we describe
later. The top abstraction was applied on two input fields of all
\code{Bet} events, namely the player position and their hand, which is
why they are not shown in Fig.~\ref{fig:tunedAutomaton}.

It is possible to apply other abstractions like the standard
numerical abstractions of abstract interpretation or a byte-length
abstraction that abstracts a value by its number of bytes, for
example, to denote the order of magnitude of assets.  For
simplicity, we focus on the abstractions that we found most useful
in our experiments.

\paragraph{Automaton moves.}
Event abstractions apply uniformly to all event fields for a given
function signature, for instance, the result values of all invocations
of \code{Bet}. It is also useful to apply abstractions more locally,
depending on the context in the automaton. We call such
automaton-specific abstractions \emph{automaton moves}.  We focus on
two important automaton moves in the following, merging states and
merging symbolic variables.

The automata constructed from abstract histories are tree-shaped and
may include long chains of states. To obtain more readable automata,
we merge states, which may join such chains and, more importantly,
introduce cycles and substantially reduce the total number of states
in an automaton.

\begin{definition}[State merge]
  Given an automaton $\mathcal{A}$ and a function $f$ from states to
  states, a \emph{state merge} is a move that produces an automaton
  $\mathcal{A}'$, where every state $q$ is replaced by $f(q)$ and
  every transition $q \overset{e}{\longmapsto} r$ in $\mathcal{A}$ is
  replaced by $f(q) \overset{e}{\longmapsto} f(r)$ in $\mathcal{A}'$.
\end{definition}

Based on this definition, we present two instantiations of function
$f$, which are based on the future of automaton states and are used by
our technique:
\begin{enumerate}
\item[--] \emph{Same bounded future}: Given a state $q$ in an automaton
  $\mathcal{A}$, $L_q$ denotes the set of words accepted by
  $\mathcal{A}$ starting from $q$. For $k \in \mathbb{N}$, $L_q^k$
  denotes the set of words that are prefixes of words in $L_q$ with
  length at most $k$. We consider two automaton states $q$ and $r$ to
  have the same bounded future if $L_q^k = L_r^k$. To merge states with the
  same bounded future, our technique selects a random value for $k$ and
  defines function $f$ such that $f(q) = f(r)$ if and only if $L_q^k =
  L_r^k$.
\item[--] \emph{Similar bounded future}: We consider two automaton states $q$ and
  $r$ to have a similar bounded future when $L_q^k \subset L_r^k$ or $L_r^k
  \subset L_q^k$. Our technique selects a random value for $k$ and
  defines function $f$ such that $f(q) = f(r)$ if and only if $L_q^k
  \subset L_r^k$ or $L_r^k \subset L_q^k$.
\end{enumerate}

Note that both variations of this automaton move potentially
make the resulting automaton very imprecise, for instance, if $k$ is
very small. However, the subsequent evaluation of the cost metric will
detect such overly coarse abstractions and reject the recipe.
In Fig.~\ref{fig:tunedAutomaton},
state~2 is the result of merging several states that led to betting in
the future.

As discussed above, an important event abstraction is to replace
concrete values by symbolic variables. It is often useful to track
relationships between different symbolic variables, especially,
the equality of variables. Since we need to relate symbolic variables
also across different histories, we cannot simply apply relational
abstractions. Instead, we define an automaton move that expresses the
equality of two symbolic variables by merging them.

\begin{definition}[Variable merge]
  Given an automaton, a \emph{variable merge} is an automaton move
  that:
  \begin{enumerate}[a.]
    \item selects two distinct symbolic variables, $v_1$ and $v_2$, that appear
      in event fields of the automaton;
    \item replaces every occurrence of $v_2$ by $v_1$;
    \item replaces $v_1$ by $*v_1$ in event fields where the variable
is assigned a different value, according to the concrete histories.
  \end{enumerate}
\end{definition}

The last step is necessary to ensure that each history of concrete
events is still
characterized by the corresponding path in the automaton.
In our example, this move was repeatedly applied
to yield the automaton of Fig.~\ref{fig:tunedAutomaton}, in which all
\code{StartGame} events return a unique game identifier and the
subsequent \code{Bet} and \code{Claim} events refer to the same
identifier.

%% ----------------------------------------------------------------
\section{Experimental Evaluation}
\label{sect:Experiments}
%% ----------------------------------------------------------------

So far, we have shown the usability of our technique on the
rock-paper-scissors contract. Here, we further evaluate its
practicality on several real-world contracts deployed on the
Ethereum blockchain.

\paragraph{Setup.}
For our experiments, we selected nine popular contracts from five
categories that represent common applications of smart contracts. In
particular, we chose contracts that implement auctions, gambling
games, Ponzi schemes~\cite{BartolettiCarta2017},
tokens\footnote{Tokens allow clients to create and manage their own
  crypto-currency.}, and wallets. An overview of these contracts is
shown in the first column of Tab.~\ref{tab:Histories}, grouped by the
above categories (in the same order). The second column shows the
lines of code for each contract, where available.

For each contract, we identified an interesting seed transaction as
input to our tool, in particular, a transaction that instantiates the
contract or one that marks the beginning of interesting interactions,
like the start of an auction.
%
%% (We provide the exact seed transactions
%% for all contracts in the supplementary material.)
%
For all experiments, we set an exploration bound of 10,000 recipes.

We ran the experiments on an Intel Xeon CPU E5-4627 v2 @ 3.30GHz
machine with 256GB of memory running the Ubuntu operating system with
Linux 4.4 kernel.

\begin{table}[t!]
\centering
\caption{Results for history mining.}
%\vspace{-1em}
\scalebox{0.8}{
\begin{tabular}{c|c|c|c|c|c}
\multirow{3}{*}{\textsc{\textbf{Contract}}} & \multirow{3}{*}{\textsc{\textbf{LOC}}} & \multicolumn{2}{c|}{\textsc{\textbf{Transactions}}} & \multicolumn{2}{c}{\textsc{\textbf{Histories}}}\\
& & \multirow{2}{*}{\textsc{Total}} & \multirow{2}{*}{\textsc{Final}} & \multirow{2}{*}{\textsc{Total}} & \textsc{Average}\\
& & & & & \textsc{Length}\\
\hline
\code{ENS} & 606 & 1,859,823 & 274 & 60 & 10\\
\hline
\code{Esports} & 129 & 767,655 & 54 & 16 & 16\\
\code{Etherdice} & 976 & 708,244 & 664 & 80 & 134\\
\hline
\code{PiggyBank} & 89 & 628,585 & 150 & 123 & 71\\
\code{PonziKing} & -- & 1,163,028 & 45 & 7 & 14\\
\hline
\code{BAT} & 175 & 26,276 & 69 & 30 & 61\\
\code{REP} & -- & 21,101 & 65 & 3 & 23\\
\code{TheDao} & 1,236 & 3,535 & 188 & 33 & 81\\
\hline
\code{EthDev} & -- & 603,106 & 105 & 147 & 38\\
\end{tabular}}
\label{tab:Histories}
\vspace{0em}
\end{table}

\paragraph{History mining.}
In Tab.~\ref{tab:Histories}, we evaluate the history mining component
of our tool chain. As shown in the table, the number of final
transactions that remain after filtering the dependency graphs (fourth
column) is only a small fraction of the total number of collected
transactions when starting from the seed (third column). This shows
the effectiveness of our technique in eliminating irrelevant
transactions and, thereby, obtaining precise specifications. At the
same time, the filtering retains sufficiently many transactions for
our tool to generate several histories (fifth column) of reasonable
average length (sixth column). On the other hand, the number of
histories is still manageable even though there are histories that
differ from each other only with respect to the order of their events.

As expected, the running time of the history mining is proportional to
the time it takes to execute the transactions on our copy of the
blockchain for collecting their effects. For the contracts shown in
Tab.~\ref{tab:Histories}, this time ranges from a few seconds (e.g.,
11s for \code{TheDao}) to several hours (e.g., 36h for
\code{ENS}). On average, 50\% of the total mining time is spent
on executing the transactions and collecting their read and write
effects, the other 50\% on building the dependency graph,
filtering it, and generating histories. The former part becomes
more dominant as the number of transactions increases (e.g., 86\%
for \code{ENS}). As we explained earlier, it is possible to
cache the results of this step and extend them incrementally
as the blockchain grows.
% and much higher when our tool collects a small number
%of transactions (e.g., 72\% for \code{TheDao}).

\paragraph{Automaton construction and tuning.}
In our experiments, we used three user configurations: one that
strikes a good balance between readability and precision (called the
\emph{default configuration}), one that favors generality (\emph{general
configuration}), and one that favors precision (\emph{precise configuration}).

In Tab.~\ref{tab:Automata}, we evaluate the automaton construction and
abstraction tuning component of our tool chain. Columns 2--5 of the
table, labeled `Transitions', show the number of transitions in the
initial, default, general, and precise automata. As expected, the
initial automata are too large to be readable by a user (with up to
464 transitions for \code{Etherdice}), whereas the automata generated
by the default configuration of our tool contain an average of 39
transitions. We can see that the number of transitions typically varies for different
configurations since the number of transitions is one of the factors in the cost
computation. For instance, for \code{TheDao}, the general automaton contains the fewest
number of transitions and the precise automaton contains the most. However, since we also
consider other cost factors, the same does not always hold for other smart contracts.

Columns 6--8, labeled `Accepted Recipes', show the percentage of
recipes that are accepted by our optimization algorithm. Observe that,
even in the worst case, 51\% of the generated
recipes are accepted (for \code{PonziKing}) and, therefore, contributed to the exploration;
in the best case, this percentage goes up to 88.4\% for
\code{Etherdice}.
Columns 9--11, labeled `Cost Reduction', show how much the cost is reduced between the
initial and the best automaton. For most contracts, the cost reduction is high (on average
74.6\% for the default configuration), which suggests that our technique is able to
effectively tune the automata based on the given cost configuration.

\begin{table}[t!]
\centering
\caption{Results for automaton construction and tuning.}
%\vspace{-1em}
\scalebox{0.74}{
\begin{tabular}{c|c|c|c|c|c|c|c|c|c|c}
\multirow{2}{*}{\textsc{\textbf{Contract}}} & \multicolumn{4}{c|}{\textsc{\textbf{Transitions}}} & \multicolumn{3}{c|}{\textsc{\textbf{Accepted Recipes}}} & \multicolumn{3}{c}{\textsc{\textbf{Cost Reduction}}}\\
& \textsc{I} & \textsc{D} & \textsc{G} & \textsc{P} & \textsc{D} & \textsc{G} & \textsc{P} & \textsc{D} & \textsc{G} & \textsc{P}\\
\hline
\code{ENS} & 87 & 20 & 20 & 20 & 74.7\% & 65.7\% & 83.4\% & 81.8\% & 86.1\% & 66.5\%\\
\hline
\code{Esports} & 65 & 21 & 20 & 32 & 73.3\% & 68.9\% & 82.5\% & 65.4\% & 83.7\% & 62.9\%\\
\code{Etherdice} & 463 & 37 & 81 & 460 & 78.0\% & 55.8\% & 86.6\% & 88.4\% & 91.3\% & 68.0\%\\
\hline
\code{PiggyBank} & 121 & 27 & 30 & 34 & 71.6\% & 60.0\% & 82.6\% & 77.2\% & 87.4\% & 49.3\%\\
\code{PonziKing} & 24 & 24 & 24 & 24 & 68.6\% & 51.0\% & 82.3\% & 30.7\% & 10.0\% & 58.4\%\\
\hline
\code{BAT} & 253 & 54 & 39 & 52 & 73.6\% & 74.8\% & 82.3\% & 86.8\% & 92.9\% & 59.5\%\\
\code{REP} & 66 & 20 & 35 & 35 & 53.9\% & 83.3\% & 75.7\% & 80.4\% & 73.5\% & 70.5\%\\
\code{TheDao} & 456 & 84 & 47 & 99 & 70.5\% & 69.2\% & 77.2\% & 78.5\% & 88.5\% & 77.7\%\\
\hline
\code{EthDev} & 121 & 67 & 46 & 21 & 80.8\% & 75.5\% & 61.2\% & 82.4\% & 82.1\% & 91.2\%\\
\end{tabular}}
\label{tab:Automata}
\vspace{0em}
\end{table}

With a bound of 10,000 recipes, the running time of this tool component
is between 7secs (for \code{PonziKing}) and 551mins (for
\code{Etherdice}). The latter is, however, an outlier, which we attribute
to the large number of histories and their average length
(Tab.~\ref{tab:Histories}). The final automata for the majority of
contracts and configurations are generated within less than ten minutes,
despite the high exploration bound we selected. We observed
that, even with a much smaller bound, the results are often
comparable and generated within only a few seconds for most
contracts. In general, most of the running time of this component is
spent on computing a cost for each candidate automaton.

To illustrate the effectiveness of our approach, we
describe our experience from generating the general automaton
for the \code{ENS} auction. Even though the mining process
starts out with almost two million transactions, it produces a
concise automaton with only 20 transitions. This automaton correctly
shows that auctions are started with an
invocation of \code{startAuctions}. Bids may be placed by calling
\code{newBid} or \code{startAuctionsAndBid}, which can start an
auction and place a bid at the same time. Bids are then unsealed by
invoking \code{unsealBid}. The automaton shows also that, after an
invocation of \code{unsealBid}, no more bids are placed which ensures fairness of the auction. After
the bids are unsealed, the auction may be finalized by calling
\code{finalizeAuction}. We believe the automaton provides a convenient way to extract
and visualize such information even if there are hundreds of different transactions.

\section{Related Work}

\paragraph{Specification mining.}
The problem of specification mining is fundamental and well studied in
the literature~\cite{RobillardBodden2013}. To the best of our
knowledge, this work is the first to apply specification mining in the
context of smart contracts. Early work on specification
mining~\cite{BiermannFeldman1972} phrases the problem as learning
finite-state machines from sets of input/output pairs that partially
describe their behavior. This early work presents the $k$-tails
heuristic, which merges states in order to generalize from the given
examples. In our setting, this heuristic corresponds to the
same-future state merge.

There are approaches that aim to simplify mining by restricting what
constitutes an automaton
state~\cite{DallmeierLindig2006,PradelGross2009} or by defining
discoverable patterns a
priori~\cite{GabelSu2008-Javert,GabelSu2008-Symbolic,GabelSu2010},
like the alternating pattern $(ab)^*$ (e.g., locking and unlocking
resources) or the resource usage pattern $ab^*c$ (e.g., opening,
reading, and closing files). Although scalable, these approaches can miss
interesting interactions between states.

More generally, work on specification mining may be classified into
dynamic and static approaches. Like our technique, dynamic
specification mining relies on having run the target program with
reasonable coverage. Applications of dynamic mining include software
revision histories~\cite{LivshitsZimmermann2005}, heap properties of
object-oriented programs~\cite{DemskyRinard2002}, component
interactions~\cite{MarianiPezze2007}, library
APIs~\cite{KrkaBrun2014}, system logs~\cite{BeschastnikhBrun2013},
scenario-based system
behaviors~\cite{LoMaoz2007,LoMaoz2008,LoMaoz2010,FahlandLo2013}, etc.

Static approaches are further subdivided into component- and
client-side mining. In component-side techniques, a specification is
mined by analyzing a component's
implementation~\cite{AlurCerny2005,NandaGrothoff2005}, whereas
client-side techniques generate a specification that reflects usage
patterns in a
code-base~\cite{EnglerChen2001,MandelinXu2005,WeimerNecula2005,WasylkowskiZeller2007,ShohamYahav2008,NguyenNguyen2009}.
The two kinds of approaches have been shown to complement each
other~\cite{WhaleyMartin2002}.

Specification mining approaches may also be classified into
automaton-based (e.g., \cite{CookWolf1998,LorenzoliMariani2007}) and
non-automaton-based ones (e.g.,
\cite{LieChou2001,ElRamlyStroulia2002,RazKoopman2002,LamRinard2003,YangEvans2006,ErnstPerkins2007}).
Process mining~\cite{VanDerAalstVanDongen2003} is a
non-automaton-based technique, which dynamically records system events
and mines workflow graphs. These graphs, which are similar to
petri-nets, typically encode interactions of concurrent processes.

In contrast to existing work, our technique obtains precise
specifications with a novel dependency analysis that allows us to
extract sequential and self-contained traces from the blockchain. It
also gives users the flexibility to adjust the generated automata
according to their specific needs. To achieve this flexibility, we
phrase specification mining as an optimization problem by computing a
cost for each generated automaton and automatically tuning its
abstractions. We also introduce a novel abstraction that allows us to
capture relations between values occurring in different automaton
events and fields, like the value of the game identifier in the
running example.
Our technique could easily be extended with additional metrics. For
instance, there could be scenarios where it is useful to reduce rare
state transitions by increasing their
cost~\cite{AmmonsBodik2002,LoKhoo2006}.

\paragraph{Simulated annealing.}
The algorithm for simulated annealing was first described by
Metropolis et al.~\cite{MetropolisRosenbluth1953}, but it was better
detailed many years later~\cite{ChibGreenberg1995}. Although simulated
annealing has been applied to optimization problems for a few decades
already~\cite{KirkpatrickGelatt1983}, it only recently started being
integrated with program analysis techniques (e.g.,
\cite{SharmaAiken2014,FuSu2016,FuSu2017}).
We use simulated annealing
for automatically tuning automata that describe the functionality of
smart contracts.

\paragraph{Program analysis for smart contracts.}
Smart contract code is susceptible to bugs just like any other
program, with the additional hazard of losing
crypto-assets~\cite{AtzeiBartoletti2017}. As a consequence, the
program-analysis and verification community has already developed
several bug-finding techniques for smart contracts, including
debugging, static analysis, symbolic execution, and
verification~\cite{Remix,Securify,LuuChu2016,BhargavanDelignat-Lavaud2016,ChenLi2017,ChatterjeeGoharshady2018,AmaniBegel2018,GrossmanAbraham2018,KalraGoel2018,NikolicKolluri2018}.
As mentioned earlier, analyzing smart contracts poses further
challenges due to their execution model. For instance, users have no
direct control over the order in which transactions are processed by
the miners, making smart contracts susceptible to concurrency
bugs~\cite{SergeyHobor2017}. These characteristics had to be
considered when designing our specification mining technique (e.g., by
exploring all possible orderings of transactions when computing
sessions).

\section{Conclusion}

We presented the first specification mining technique for smart
contracts. Our precise specifications provide important insights into
the functionality and interactions between smart contracts, and are
useful to develop, understand, and audit smart contracts.
Unlike existing work, our technique gives users the flexibility to
adjust the generated automata based on their needs during specific usage
scenarios. We achieve this by phrasing the problem of finding useful
abstractions for automaton construction as an optimization problem
with user-adjustable costs.

As future work, we plan to use our technique to identify common design
patterns and evolve the language design of smart contracts. We also
plan to apply the idea of cost-guided abstraction tuning to other forms of
program analysis, such as heap analyses, invariant inference, and
process mining.

%% \todo{Thank Rose and reviewers.}

%% Acknowledgments
% \begin{acks}                            %% acks environment is optional
%                                         %% contents suppressed with 'anonymous'
%   %% Commands \grantsponsor{<sponsorID>}{<name>}{<url>} and
%   %% \grantnum[<url>]{<sponsorID>}{<number>} should be used to
%   %% acknowledge financial support and will be used by metadata
%   %% extraction tools.
%   This material is based upon work supported by the
%   \grantsponsor{GS100000001}{National Science
%     Foundation}{http://dx.doi.org/10.13039/100000001} under Grant
%   No.~\grantnum{GS100000001}{nnnnnnn} and Grant
%   No.~\grantnum{GS100000001}{mmmmmmm}.  Any opinions, findings, and
%   conclusions or recommendations expressed in this material are those
%   of the author and do not necessarily reflect the views of the
%   National Science Foundation.
% \end{acks}

%% Bibliography
\bibliography{tandem}

%% Appendix
%% \appendix
%% \section{Appendix}

%% Text of appendix \ldots

\end{document}

%% file: solidity-highlighting.tex
\usepackage{listings, xcolor}

\definecolor{verylightgray}{rgb}{.97,.97,.97}

\lstdefinelanguage{Solidity}{
	keywords=[1]{anonymous, assembly, assert, balance, break, call, callcode, case, catch, class, constant, continue, contract, debugger, default, delegatecall, delete, do, else, event, export, external, false, finally, for, function, gas, if, implements, import, in, indexed, instanceof, interface, internal, is, length, library, log0, log1, log2, log3, log4, memory, modifier, new, payable, pragma, private, protected, public, pure, push, require, return, returns, revert, selfdestruct, send, storage, struct, suicide, super, switch, then, this, throw, transfer, true, try, typeof, using, value, view, while, with, addmod, ecrecover, keccak256, mulmod, ripemd160, sha256, sha3}, % generic keywords including crypto operations
	keywordstyle=[1]\color{blue}\bfseries,
	keywords=[2]{address, bool, byte, bytes, bytes1, bytes2, bytes3, bytes4, bytes5, bytes6, bytes7, bytes8, bytes9, bytes10, bytes11, bytes12, bytes13, bytes14, bytes15, bytes16, bytes17, bytes18, bytes19, bytes20, bytes21, bytes22, bytes23, bytes24, bytes25, bytes26, bytes27, bytes28, bytes29, bytes30, bytes31, bytes32, enum, int, int8, int16, int24, int32, int40, int48, int56, int64, int72, int80, int88, int96, int104, int112, int120, int128, int136, int144, int152, int160, int168, int176, int184, int192, int200, int208, int216, int224, int232, int240, int248, int256, mapping, string, uint, uint8, uint16, uint24, uint32, uint40, uint48, uint56, uint64, uint72, uint80, uint88, uint96, uint104, uint112, uint120, uint128, uint136, uint144, uint152, uint160, uint168, uint176, uint184, uint192, uint200, uint208, uint216, uint224, uint232, uint240, uint248, uint256, var, void, ether, finney, szabo, wei, days, hours, minutes, seconds, weeks, years},	% types; money and time units
	keywordstyle=[2]\color{teal}\bfseries,
	keywords=[3]{block, blockhash, coinbase, difficulty, gaslimit, number, timestamp, msg, data, gas, sender, sig, value, now, tx, gasprice, origin},	% environment variables
	keywordstyle=[3]\color{violet}\bfseries,
	identifierstyle=\color{black},
	sensitive=false,
	comment=[l]{//},
	morecomment=[s]{/*}{*/},
	commentstyle=\color{gray}\ttfamily,
	stringstyle=\color{red}\ttfamily,
	morestring=[b]',
	morestring=[b]"
}